\documentclass[conference]{IEEEtran}
\IEEEoverridecommandlockouts
\usepackage{cite}
\usepackage{amsmath,amssymb,amsfonts}
\usepackage{algorithmic}
\usepackage{graphicx}
\usepackage{textcomp}
\usepackage{xcolor}

\usepackage{booktabs}
\usepackage{listings}
\usepackage{multicol}
\usepackage{multirow}
\usepackage{makecell}
\usepackage{subfigure}
\usepackage[hidelinks]{hyperref}
\def\BibTeX{{\rm B\kern-.05em{\sc i\kern-.025em b}\kern-.08em
    T\kern-.1667em\lower.7ex\hbox{E}\kern-.125emX}}
\begin{document}

\title{Large Language Models for Anomaly Detection in Computational Workflows: from Supervised Fine-Tuning to In-Context Learning}

\author{
  \IEEEauthorblockN{
    Hongwei Jin\IEEEauthorrefmark{1},
    George Papadimitriou\IEEEauthorrefmark{2},
    Krishnan Raghavan\IEEEauthorrefmark{1},
    Pawel Zuk\IEEEauthorrefmark{2},\\
    Prasanna Balaprakash\IEEEauthorrefmark{3}
    Cong Wang\IEEEauthorrefmark{4},
    Anirban Mandal\IEEEauthorrefmark{4},
    Ewa Deelman\IEEEauthorrefmark{2},
  }
  \IEEEauthorblockA{
    \IEEEauthorrefmark{1}Mathematics and Computer Science Division, Argonne National Laboratory, Lemont, USA\\
    \{jinh, kraghavan\}@anl.gov
  }
  \IEEEauthorblockA{
    \IEEEauthorrefmark{2}Information Sciences Institute, University of Southern California, Los Angeles, USA\\
    \{georgpap, pawelzuk, deelman\}@isi.edu}
  \IEEEauthorblockA{
    \IEEEauthorrefmark{3}Computing and Computational Sciences Directorate , Oak Ridge National Laboratory, Oak Ridge, USA\\
    pbalapra@ornl.gov}
  \IEEEauthorblockA{
    \IEEEauthorrefmark{4}RENCI, University of North Carolina at Chapel Hill, Chapel Hill, USA\\
    \{cwang, anirban\}@renci.org}
}
\maketitle

\begin{abstract}
  %
  Anomaly detection in computational workflows is critical for ensuring system reliability and security. However, traditional rule-based methods struggle to detect novel anomalies. This paper leverages large language models (LLMs) for workflow anomaly detection by exploiting their ability to learn complex data patterns. Two approaches are investigated: 1) supervised fine-tuning (SFT), where pre-trained LLMs are fine-tuned on labeled data for sentence classification to identify anomalies, and 2) in-context learning (ICL) where prompts containing task descriptions and examples guide LLMs in few-shot anomaly detection without fine-tuning. The paper evaluates the performance, efficiency, generalization of SFT models, and explores zero-shot and few-shot ICL prompts and interpretability enhancement via chain-of-thought prompting. Experiments across multiple workflow datasets demonstrate the promising potential of LLMs for effective anomaly detection in complex executions.
\end{abstract}


\section{Introduction}
\label{sec:intro}
With the increasing complexity and scale of modern systems, computational workflows are growing in complexity while their reliability, security, and performance are becoming rather important. A critical factor in ensuring workflow execution reliability is the ability to detect anomalies. These anomalies can be indicators of various system issues, and they are manifested by unexpected behavior in hardware, such as high usage of computing resources, memory consumption, and I/O operations. To address the problem of anomaly detection in modern systems, methods that rely on rule-based systems, statistical analysis, and machine learning techniques \cite{borghesi2019anomaly, kiran2020detecting, herath2019ramp, jin2022workflow} have become quite popular in recent years.

Despite their effectiveness, a considerable amount of data preprocessing must be done to perform this detection because typical methods are limited to analyzing images or numerical values. Furthermore, to facilitate this data preprocessing, a lot of expert knowledge is needed to be put into carefully collecting and correlating low-level system statistics with workflow execution metadata that can be used to convert the raw logs into other formats.
Adding to the complexity is the need for substantial ML expertise to navigate the wide array of available anomaly detection methodologies effectively. The field of ML presents a vast spectrum of models and techniques, each with its customization and application nuances. This diversity, while beneficial, also imposes a steep learning curve and necessitates a deep understanding of ML principles to tailor these models to specific anomaly detection tasks. Furthermore, the process of setting up and training these models—integrating them into a system's workflow—poses an additional challenge. This aspect of ML model deployment and maintenance may not align well with the skill set of system administrators, who are typically more versed in direct system maintenance rather than in the nuances of ML model training and tuning.

Large Language Models (LLMs) and their wide-spread democratization efforts have the potential to significantly transform anomaly detection in HPC systems by streamlining data preprocessing, enhancing pattern recognition, simplifying the deployment of machine learning models, enabling real-time monitoring, and fostering a supportive community ecosystem. By automating complex data processing tasks and offering advanced analytical capabilities, LLMs reduce the need for extensive expert knowledge, making sophisticated anomaly detection accessible to system administrators without deep technical backgrounds. Furthermore, their ability to process and analyze streaming data in real-time can ensure prompt detection and mitigation of potential system issues.

A primary critique of LLMs concerns their energy/power consumption and model size, which are seen as barriers to their practical application in HPC data analysis. However, this perspective overlooks the significant advances in energy-efficient technologies and the optimization of LLMs for operation on a wide range of devices, from high-end servers to compact, low-power devices such as smartphones. These emerging technologies not only mitigate the energy and resource demands of running sophisticated LLMs but also expand their accessibility and usability across various platforms. Consequently, as these energy-efficient techniques continue to evolve and LLMs become increasingly optimized for smaller devices, the practicality of deploying LLMs for anomaly detection in HPC systems—and beyond—becomes ever more feasible. This trajectory underscores the viability of LLMs as a transformative tool in anomaly detection, promising significant advancements in HPC system management and maintenance.


We develop an approach that leverages pre-trained Large Language Models (LLMs) to directly detect anomalies from log files generated during the execution of computational workflows. Specifically, we adapt these pre-trained models through Supervised Fine-Tuning (SFT) and Prompt Engineering via In-Context Learning (ICL).
SFT employs a pre-trained model and trains on a smaller dataset of labeled examples for a specific task~\cite{ouyang2022training}. Unlike the training of LLMs in an unsupervised way, the SFT often consists of an input and a desired output. By updating the parameters of LLMs again through SFT, the model improves the performance for a downstream task. However, one common issue with LLMs is that they can perpetuate biases present in the data used to train them~\cite{wang2023overwriting}, especially when for the binary classification problem. Another common issue is catastrophic forgetting~(CF)~\cite{luo2023empirical}, which occurs in machine learning when a model forgets previously learned information as it learns new information. This is a common problem in supervised fine-tuned models, where the model is trained on a new task after it has already been trained on one or more previous tasks.

In-context learning (ICL), on the other hand, is an emerging paradigm where LLMs perform tasks by leveraging a few examples provided within the context of query~\cite{brown2020language} rather than relying on supervised fine-tuning with labeled data.
ICL heavily relies on prompt engineering, providing examples and contextual cues that guide the LLMs in efficiently understanding and executing the desired task.
A well-engineered prompt not only presents the LLM with relevant information but subtly instructs it on generating the appropriate output. It involves structuring the examples in a way that highlights patterns or relationships, using natural language templates that align with the task's goals, or including explicit instructions that direct the model's attention to critical aspects of the problem. This alignment can be performed specifically for the anomaly detection problem where the prompts contain information about the job features and brief statistics about the job execution facilitating anomaly detection in the workflow.
Furthermore, prompts can also include the instruction for reasoning steps through Chain-of-Thought (CoT~\cite{wei2022chain}), providing explainable output from LLMs.


\begin{figure*}[th]
  \centering
  \includegraphics[width=0.8\linewidth]{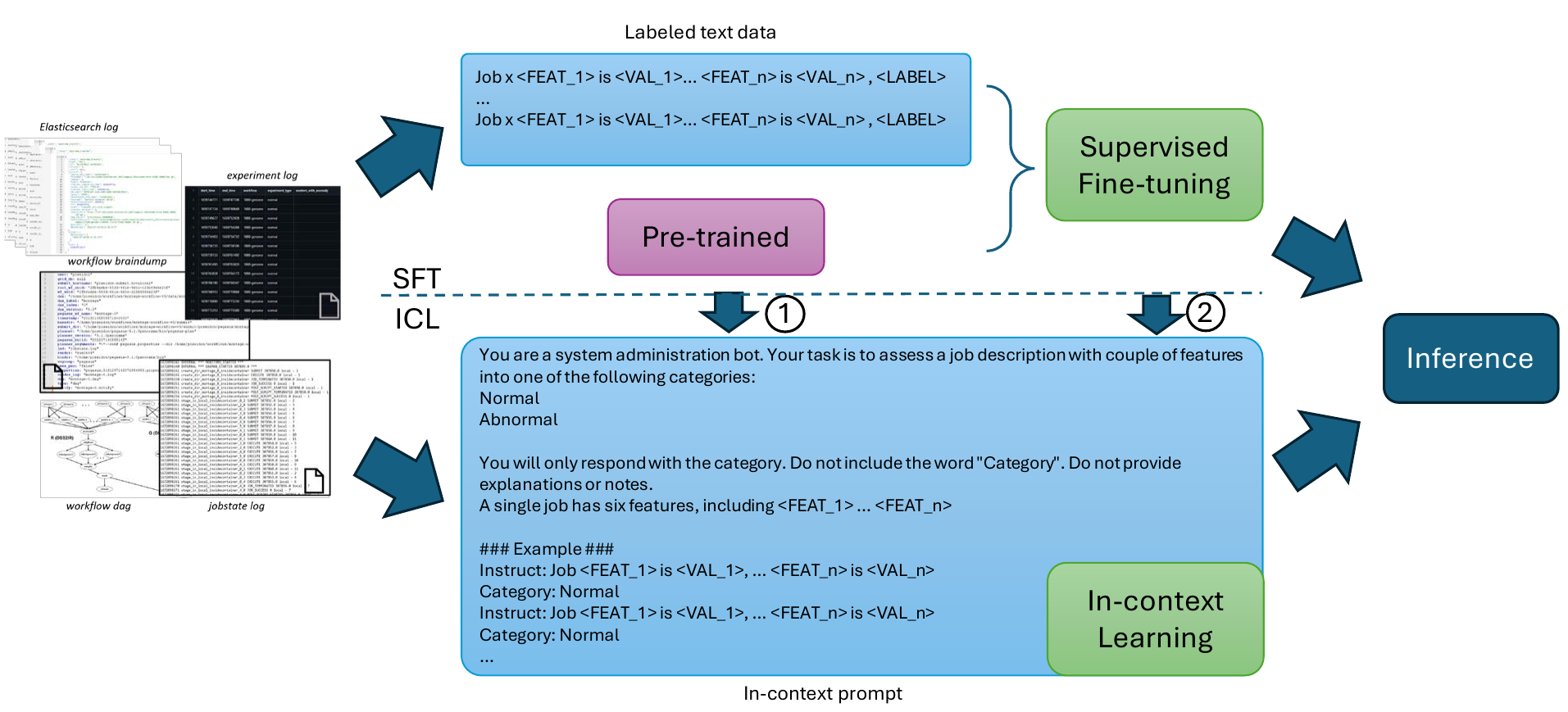}
  \caption{Supervised Fine-tuning and In-Context Learning for anomaly detection}
  \vspace{-1em}
  \label{fig:sft_diagram}
\end{figure*}


To this end, we make the following contributions to the scope of this paper:
\begin{enumerate}
  \item Investigate the efficacy of LLMs for anomaly detection and evaluate the performance of supervised fine-tuning models and in-context learning in detecting anomalies in computational workflows.
  \item Address the biases, overcome the catastrophic forgetting, and explore the generalization through transfer learning and online detection.
  \item Explore the ability of ICL with zero-shot, few-shot learning, and study the interpretable output from ICL through Chain-of-Thought~(CoT).
\end{enumerate}
With the use of LLMs for anomaly detection in computational workflows, we seek to contribute to the development of effective and efficient methods for detecting anomalies in computational workflows.

\vspace{-.5em}

\section{Background and Related Work}
\label{sec:background}

Several approaches have been proposed in the literature for anomaly detection in computational workflows.
These approaches can be broadly classified into rule-based systems, statistical analysis, and machine-learning techniques. Rule-based systems rely on predefined thresholds and patterns to detect anomalies.
For example, \cite{thompson2015detecting} proposed a rule-based system that uses a set of heuristics to identify anomalies in Linux computational workflows.

While rule-based systems are simple to implement, they are limited by their inability to adapt to changing patterns in behavior and are often brittle in the face of new anomalies. Statistical analysis techniques, that use statistical information such as mean, median, and standard deviation, have been used to detect anomalies in computational workflows as well.
For example, \cite{chandola2009anomaly} proposed an approach that uses statistical methods to identify anomalies in network traffic logs. However, statistical analysis techniques are sensitive to outliers and may not be effective in detecting anomalies that do not deviate significantly from the mean.

Machine learning techniques, such as decision trees, random forests, and clustering, have also been applied to anomaly detection in computational workflows.
For example, \cite{ng2001spectral} presented a simple and effective algorithm for spectral clustering, a method for grouping data points based on their pairwise similarities. The algorithm utilizes the eigenvectors of the graph Laplacian to represent similarities between the data points. The authors also provide a theoretical analysis of the algorithm and show that it can be used to cluster data in a variety of settings.

More recently, deep learning techniques, such as recurrent neural networks (RNNs) and convolutional neural networks (CNNs), have been applied to anomaly detection in computational workflows.
For example, \cite{xu2009detecting} proposed a general methodology for mining console logs to detect large-scale system problems. The authors first parse the logs by combining source code analysis with information retrieval to create composite features. They then analyze these features using machine learning to detect operational problems. The authors evaluate their methodology on a dataset of console logs from a large-scale production system and show that it can be used to effectively detect a variety of system problems, including performance problems, software bugs, and malicious activity.

Several works also take advantage of LLMs for anomaly detection in system logs.
LogBERT \cite{liu2021logbert} is a self-supervised anomaly detection framework based on BERT that learns the patterns of normal log sequences by two novel self-supervised training tasks and is able to detect anomalies where the underlying patterns deviate from normal log sequences. Similarly, BERT-log \cite{wang2022bertlog} also trains a BERT model but with labeled data to detect anomalies in logs.
UniLog \cite{zhu2021unilog} and LTanomaly \cite{xu2022ltanomaly} are both Transformer-based anomaly detection methods for system logs, but UniLog is a pre-trained model, while LTanomaly is a Transformer variant that is specifically designed for syslog anomaly detection. While impressive results have been demonstrated, these approaches are not easily extensible and applicable to other workflows beyond the ones used in these papers. This is because these papers introduce their own unique tokenization which does not generalize to different logging systems with different vocabularies. This limits the usage of these approaches once a new logging system is deployed.
In contrast, our approach leverages pre-trained models, and therefore, is easy to generalize to different kinds of logs and different workflows, which is demonstrated in our results.

\section{LLMs for Anomaly Detection}
\label{sec:llm_ad}
In this section, we will describe the supervised fine-tuning and in-context learning in details, and their advantages in anomaly detection tasks. An overview of our approach is provided in Figure~\ref{fig:sft_diagram}.
\vspace{-.5em}
\subsection{Supervised Fine-Tuning}
Supervised fine-tuning (SFT) is used to adapt pre-trained language models to new tasks or domains. The process involves feeding a labeled dataset of the target task or domain to the pre-trained model and adjusting the model's parameters while minimizing the loss on the new task. By using labeled data from the target task, the model can learn to recognize patterns and features that are specific to the new task, while still leveraging the knowledge it has gained from the large amounts of data it was pre-trained on.

Following this, we detect the anomalies in computational workflows by fine-tuning the pre-trained models on the labeled dataset of the target task, i.e., sentence classification. Our approach involves treating the logs generated by the computational workflows as a sequence of sentences and applying the fine-tuned model to classify each sentence as normal or anomalous. Toward this end, we use a combination of pre-trained models and evaluate their performance on Flow-Bench dataset. A template that parses a system log entry into a sentence with labels is provided in Figure~\ref{fig:parsed_log_template}.

Instead of training LLMs from scratch as done in \cite{wang2022bertlog, guo2021logbert}, there are several advantages of using the SFT approach:
\begin{itemize}
  \item \textbf{Reduced training time and resources}: SFT allows us to leverage the knowledge gained by the pre-trained model, reducing the amount of training time and resources required to achieve good performance on the target task. This can save a significant amount of time and computational resources.
  \item \textbf{Improved performance}: SFT has been shown to improve the performance of pre-trained models on a wide range of NLP tasks, including text classification, sentiment analysis, and question answering. By adapting pre-trained models to the target task, we can achieve better performance than training a model from scratch.
  \item \textbf{Easy domain adaptation}: SFT allows us to adapt pre-trained models to new domains, enabling them to learn domain-specific features and patterns. This can be useful for tasks like anomaly detection, where the target domain may be different from the domain the model was pre-trained on.
  \item \textbf{Better Generalization}: SFT can lead to better generalization to unseen data compared to training a model from scratch since the pre-trained model has already learned to recognize many features that are useful for the target task.
  \item \textbf{Smaller dataset requirements}: SFT can be more effective with smaller datasets than training a model from scratch since the pre-trained model has already learned to recognize many features that are useful for the target task. This can be particularly useful for tasks where labeled data is scarce or difficult to obtain, e.g., anomalies in computational workflows.
\end{itemize}
\begin{figure}
  \centering
  \begin{lstlisting}
    <FEAT_1> is <VAL_1> <FEAT_2> is <VAL_2> ... 
    <FEAT_n> is <VAL_n>, <LABEL>
  \end{lstlisting}
  \caption{Template of parsed log into a sentence.}
  \vspace{-1em}
  \label{fig:parsed_log_template}
\end{figure}
\subsection{In-Context Learning}
In-context learning (ICL) explores the LLMs' ability to enable few-shot learning and improve the generalization capabilities of the model. In contrast to the SFT, ICL does not train the model explicitly, instead, it applies prompts (input context) to guide the LLMs applying on downstream tasks. To highlight the ICL approach, we highlight several advantages of using ICL as follows:
\begin{itemize}
  \item \textbf{Improved generalization}: ICL enables models to learn from the context provided in the input data, which can improve their generalization capabilities. This can be especially useful for anomaly detection in system logs, where the data can be highly variable and complex.
  \item \textbf{Reduced need for labeled data}: ICL can enable models to learn from unlabeled data, reducing the need for expensive and time-consuming labeling efforts. This can be particularly beneficial in the context of anomaly detection, where labeling data can be difficult and resource-intensive.
  \item \textbf{Improved interpretability}: ICL also provides insights into the features and patterns that are important for detecting anomalies, making it easier to interpret the model's predictions and identify false positives or negatives. Especially, Chain-of-Thought (CoT)~\cite{wei2022chain} is a method that can be used to generate prompts that guide the model to generate the desired output by providing explainable results.
\end{itemize}
Under the ICL paradigms, there are different types of prompts that can be used to guide the LLMs' learning, including zero-shot prompts, one-shot prompts, and few-shot prompts. Zero-shot prompts provide the model with a natural language description of the task, without any examples. In this case, the model must rely solely on its prior knowledge and the context provided to explore the ability of LLMs. One-shot prompts and few-shot prompts provide the model with either a single example or a few examples of the task, respectively. Generally, the examples provided involve the label of cases, particularly in the anomaly detection task, the example could be either the normal, anomalous or even mixed examples together. This is useful for tasks where labeled data is scarce or difficult to obtain, as it allows the model to learn from a small amount of data.
Figure~\ref{fig:icl_template} provides the template of the prompt for ICL. It contains two parts in general, the task description, which guides the LLMs to understand the task, and the examples, which provide the context for the task. In our case, we explicitly ask the model to output the category of job described, without any reasoning or explanation. The contextual example, in this case, is the sentences that describe the job with its features extracted from the raw log file, and explicitly note the label of the job.

Besides, another key advantage of ICL is that it can be fine-tuned based on domain-specific datasets as well, enabling it to adapt to new domains and tasks.
Similar to SFT, fine-tuning on ICL also applies the labeled data from the target domain, capturing the specific features and patterns that are relevant to the task.
\begin{figure}[h]
  \centering
  \begin{lstlisting}
  # Prompt of task for ICL
  You are a system administration bot. 
  Your task is to assess a job description with a couple 
  of features into one of the following categories: 
  Normal and Abnormal

  You will only respond with the category. 
  Do not include the word "Category". 
  Do not provide explanations or notes.
  A single job includes <FEAT_1> ... <FEAT_n>

  # Example prompt
  Instruct: <FEAT_1> is <VAL_1>, ... <FEAT_n> is <VAL_n>
  Category: Normal
  \end{lstlisting}
  \caption{Template of in-context learning.}
  \vspace{-1em}
  \label{fig:icl_template}
\end{figure}

\subsection{Pre-trained Models}

Pre-trained models, such as BERT~\cite{devlin2018bert}, GPT~\cite{radford2018improving}, and ALBERT \cite{lan2019albert} leverage the Transformer architecture~\cite{vaswani2017attention} to ascertain statistical patterns and linguistic structures in the data. These models, trained on the large corpus of freely available text data have become the backbone of many state-of-the-art NLP systems, empowering researchers and practitioners to achieve remarkable performance with reduced training time and resources.
These models have accelerated progress in NLP and continue to drive advancements in various language understanding and generation tasks.

For text classification tasks, where the goal is to assign a category or label to a given text input, encoder-only models are commonly employed.
These models, such as BERT~\cite{devlin2018bert} (Bidirectional Encoder Representations from Transformers) and RoBERTa~\cite{liu2019roberta} (Robustly Optimized BERT Approach), process the input text in its entirety and generate contextualized representations, which can then be used for classification.
Typically, the SFT for classification tasks involves adding a classification head on top of the pre-trained model and fine-tuning the model on a labeled dataset.

On the other hand, for causal language modeling tasks, where the objective is to predict the next token in a sequence given the preceding context, decoder-only models are well-suited. These models, such as GPT~\cite{radford2018improving} (Generative Pre-trained Transformer) and its variants, generate text in an autoregressive manner, making them suitable for tasks like text generation, machine translation, and summarization.
Unlike the SFT, which predicts the label of the given sentences, ICL outputs more context-aware results, which involve the generation of words and sentences based on the context provided.

To the scope of our anomaly detection task, we will select a set of encoder-only models for SFT tasks, and decoder-only models for ICL tasks.

\section{Experiments}
\label{sec:experiments}

Our experiments are conducted on a single NVIDIA A100 GPU with 40GB memory. We implemented in PyTorch~\cite{paszke2019pytorch} and Huggingface's Transformers library~\cite{wolf2020transformers} for our experiments. The detailed configurations of each individual model and optimizer are presented in the Artifact Appendix.

\subsection{Dataset and Data Processing}
To conduct our experiments we adopted the workflow data from Flow-Bench~\cite{papadimitriou2023flowbench}, a collection of three computational workflows for anomaly detection. The contributors of this dataset manually created a set of anomaly templates that represent different types of anomalies that could occur in the workflow data, such as missing data, incorrect data, or unexpected patterns (e.g., performance degradation).
They then adopted these templates to create instances of anomalies in the data by injecting them into real workflow executions, at various points.
The benchmark design also took steps to ensure that the anomalies were realistic and representative of real-world scenarios.
For example, they ensured that the anomalies were not too frequent or too rare and that they were distributed across the data in a way that was consistent with real-world patterns.
Flow-Bench contains 1211 execution traces of three computational workflows, that we briefly describe here.
\begin{itemize}
  \item The \textit{1000 Genome Workflow} identifies mutational overlaps using data from the 1000 Genomes Project~\cite{1000genome-project} in order to provide a null distribution for rigorous statistical evaluation of potential disease-related mutations across populations. The instance of the workflow DAG available in the dataset, has a total of 137 jobs nodes and 289 edges.
  \item The \textit{Montage Workflow} uses the Montage astronomical image toolkit~\cite{2010ascl.soft10036J} to transform astronomical images, captured by the Digitized Sky Survey (DSS)~\cite{dss-archive}, into custom mosaics for further analysis of the deep sky. The instance of the workflow DAG available in the dataset, has a total of 539 nodes and 2838 edges.
  \item The \textit{Predict Future Sales Workflow} uses real historical sales data in order to train machine learning models that accurately predict the sales of the following month. The instance of the workflow DAG available in the dataset, has a total of 165 nodes and 581 edges.
\end{itemize}
Additionally, in Flow-Bench, alongside the normal - baseline data, the authors have included two main anomaly classes that model performance degradation, CPU and HDD, with multiple subclasses based on the magnitude of the slow downs. In the CPU case, the authors instructed their workers to advertise a fixed number of cores, but then use affinity and cgroups to limit the actual cores that could do processing. In the case of HDD, they limited the average read and write speed of their workers. For a more detailed description of the workflows and the data available in Flow-Bench we would like to redirect the reader to~\cite{papadimitriou2023flowbench}.

To process the logs, we first convert them into tabular format, where each row represents a log entry and each column represents a field in the log, including timestamps, job status, time duration, I/O operations, etc.
In order to omit the variance of timestamps, we select time durations of states of a job, along with I/O and CPU operations as the features for anomaly detection. Figure~\ref{fig:parsed_log_template} provides a template of the parsed log into sentences.

Having generated the parsed sentences for each job, we split the entire dataset into train, validation, and test sets with the ratio of 8:1:1. Table~\ref{tab:dataset_stats} shows the statistics of the dataset, including number of normal and anomalous nodes (jobs), and the percentage of anomalies in each split.
\begin{table}
  \centering
  \caption{Dataset statistics}
  \label{tab:dataset_stats}
  \resizebox*{\linewidth}{!}{
    \begin{tabular}{c|rccc}
      \toprule
      Dataset & Split      & \# of normal nodes & \# of anomalous nodes & \% of anomalies \\
      \midrule
      \multirow[]{3}{*}{1000 Genome}
              & train      & 25911              & 12558                 & 0.3264          \\
              & validation & 3258               & 1551                  & 0.3225          \\
              & test       & 3229               & 1580                  & 0.3286          \\
      \midrule
      \multirow[]{3}{*}{Montage}
              & train      & 109738             & 28246                 & 0.2047          \\
              & validation & 13735              & 3513                  & 0.2037          \\
              & test       & 13756              & 3492                  & 0.2025          \\
      \midrule
      \multirow[]{3}{*}{\begin{tabular}{@{}c@{}}Sales\\Prediction\end{tabular}}
              & train      & 58043              & 13237                 & 0.1857          \\
              & validation & 7250               & 1660                  & 0.1863          \\
              & test       & 7316               & 1594                  & 0.1789          \\
      \bottomrule
    \end{tabular}
  }
\end{table}

\subsection{SFT Models}
\begin{figure}
  \centering
  \includegraphics[width=\linewidth]{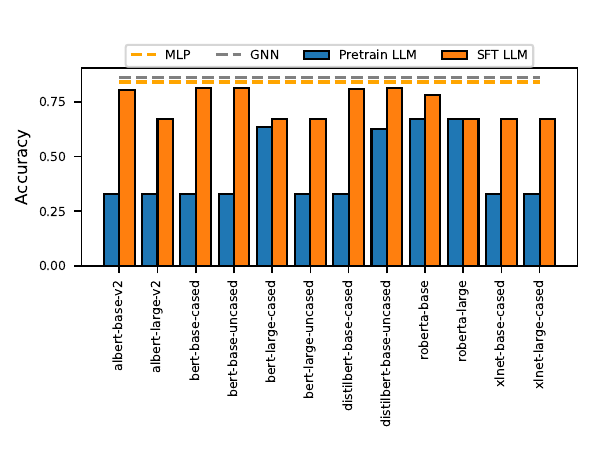}
  \caption{Reported accuracy from pre-trained models and SFT models on 1000 Genome dataset.}
  \vspace{-1em}
  \label{fig:pretrain_sft_barplot}
\end{figure}

We begin our results by validating how much we can gain from supervised fine-tuning by comparing the performance of pre-trained models and SFT models on the test set.
Figure~\ref{fig:pretrain_sft_barplot} shows the accuracy of pre-trained models and SFT models on the test set of 1000 Genome dataset. Notably, SFT models outperform pre-trained models in general, with a significant margin of improvement in several models.
In addition to the LLMs, we include conventional machine learning models, MLP and GNN, as baselines for comparison. Following the setup of models in the work~\cite{jin2023graph}, our results demonstrate that the SFT models achieve comparable performance to these classical machine learning models. However, the SFT models offer the advantage of being more versatile and accessible for the anomaly detection task, even without requiring extensive machine learning expertise.

\begin{figure}
  \centering
  \includegraphics[width=\linewidth]{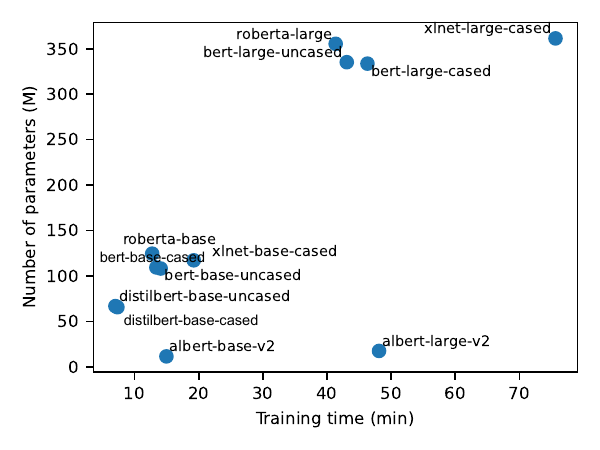}
  \vspace{-1em}
  \caption{Training time vs. number of parameters.}
  \vspace{-1em}
  \label{fig:traintime_numparams}
\end{figure}

\begin{figure}
  \centering
  \includegraphics[width=\linewidth]{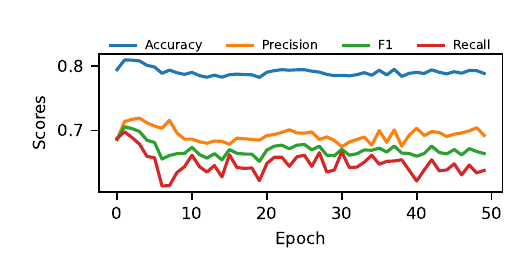}
  \caption{Validation scores with SFT.}
  \vspace{-1em}
  \label{fig:training_cost}
\end{figure}

We also try to identify the relationship between model size and its performance.
Figure~\ref{fig:traintime_numparams} shows the training time of 1000 Genome dataset and the number of parameters for the SFT models. The training time increases with the number of parameters, which is expected. However, the performance of SFT models does not necessarily increase with the number of parameters. For example, two \texttt{distilbert} models have a good performance in accuracy, while a larger model, \texttt{xlnet-base-cased} takes a longer time to train with a larger amount of parameters, resulting in worse performance when compared with \texttt{distilbert}. It is not necessary to conclude that a larger model is not helpful; the performance highly depends on the model being fine-tuned and the size of data used for training. Insufficient training data may result in underfitting in the \texttt{xlnet} model due to a limited ability to learn better embeddings for the task.
%
Lastly, we also evaluate the performance of SFT models concerning the training epochs.
As training LLMs requires a significant investment of time and resources, we aim to determine the potential benefits of increasing the number of epochs.
Figure~\ref{fig:training_cost} shows that accuracy, F1, precision, and recall scores on the validation set improve with just a few epochs of training.
However, additional epochs leads to overfitting, resulting in worse performance.
It is worth pointing out that the training time per epoch on 1000 Genome data is about 260 seconds on average.
Therefore, in practice, a few epochs of supervised fine-tuning are sufficient to transfer the model to the target task.

\subsection{Online Detection}
Online detection of anomalies in computational workflows is a critical task that can help identify potential security threats or system failures in real time.
SFT models have emerged as a powerful tool for this task, leveraging the knowledge gained from large amounts of labeled data to adapt to new tasks and domains. With the automatically parsed text sentence, we can apply the SFT models to predict the label of the system logs in real time.

\begin{figure*}
  \begin{lstlisting}
  T1: wms_delay is 6.0
     ==> label: LABEL_0, score: 0.7708
  T2: wms_delay is 6.0 queue_delay is 22.0
     ==> label: LABEL_0, score: 0.8103
  T3: wms_delay is 6.0 queue_delay is 22.0 runtime is 2090.0
     ==> label: LABEL_0, score: 0.6631
  T4: wms_delay is 6.0 queue_delay is 22.0 runtime is 2090.0 post_script_delay is 5.0
     ==> label: LABEL_1, score: 0.5780
  T5: wms_delay is 6.0 queue_delay is 22.0 runtime is 2090.0 post_script_delay is 5.0 stage_in_delay is 1310.0
     ==> label: LABEL_1, score: 0.5742
  \end{lstlisting}
  \caption{Example of online detection.}
  \vspace{-1em}
  \label{fig:online_detection}
\end{figure*}

Figure~\ref{fig:online_detection} depicts an illustration of real-time anomaly detection.
The figure shows the timestamp at T1 the computational workflows indicate that the \texttt{wms\_delay is 6.0}, which the SFT model identifies as a normal occurrence. However, as new log data becomes available, e.g., at T4, the SFT model is able to detect anomalies in real time, allowing for the prompt identification and mitigation of potential issues.

\begin{figure}
  \centering
  \includegraphics[width=0.9\linewidth]{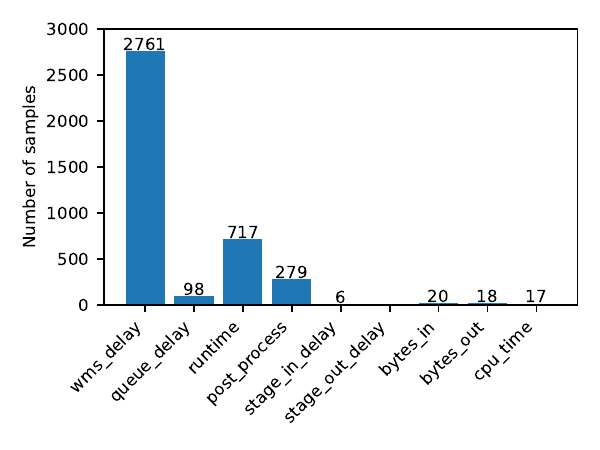}
  \caption{Early detection with online detection.}
  \vspace{-1em}
  \label{fig:early_detection}
\end{figure}

Meanwhile, we also evaluate the early detection of SFT models by checking the first time the model predicts a correct label of the job.
Figure~\ref{fig:early_detection} shows the statistics of early detection in the test set of 1000 Genome dataset, where the x-axis is the feature processed from the log in sequential order and the y-axis is the number of samples that are first identified successfully. Recall the feature of the job, involving the timestamp of each stage, we can identify the stage of the job when the anomaly is detected. For example, the anomaly is detected at the first stage of the job, which is the \texttt{wms\_delay} stage. The figure shows that the SFT models can detect anomalies at the early stage of the job, which could significantly help mitigate the potential issues of the job.

\subsection{Debiasing LLMs}

In the context of anomaly detection, the LLMs may output biased labels for normal and anomalous inputs, which can lead to incorrect or unfair results. One source of bias in SFT is the pre-trained LLM itself. LLMs are trained on massive datasets of text and code, which may contain biases that are reflected in the model's output. For example, an LLM trained on a dataset of news articles may be biased towards certain political viewpoints.

Another source of bias in SFT is the dataset utilized for fine-tuning. If the dataset is not representative of the population that the LLM will be used on, this can lead to biases in the model's output. For example, if the dataset for a text classification task only contains examples written by white males, the model may be biased against other groups of the population. Ideally, given the empty sentence, which means without any pre-knowledge of the job, the model should predict normal and abnormal jobs with almost equal probability.

\begin{figure*}
  \subfigure[Model prediction without data augmentation.]{
    \includegraphics[width=0.45\linewidth]{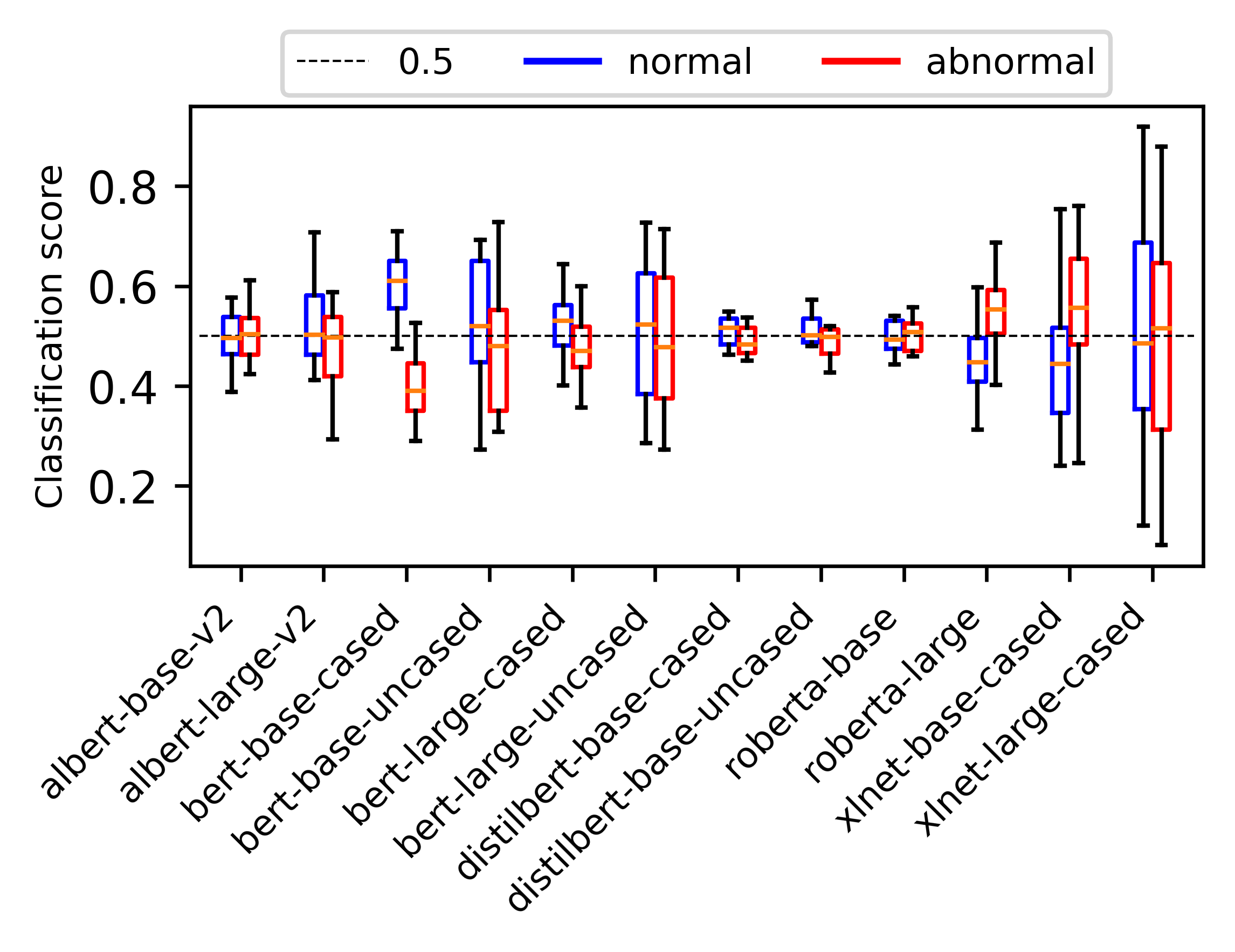}\label{fig:a}}
  \hfill
  \subfigure[Model prediction with data augmentation.]{
    \includegraphics[width=0.45\linewidth]{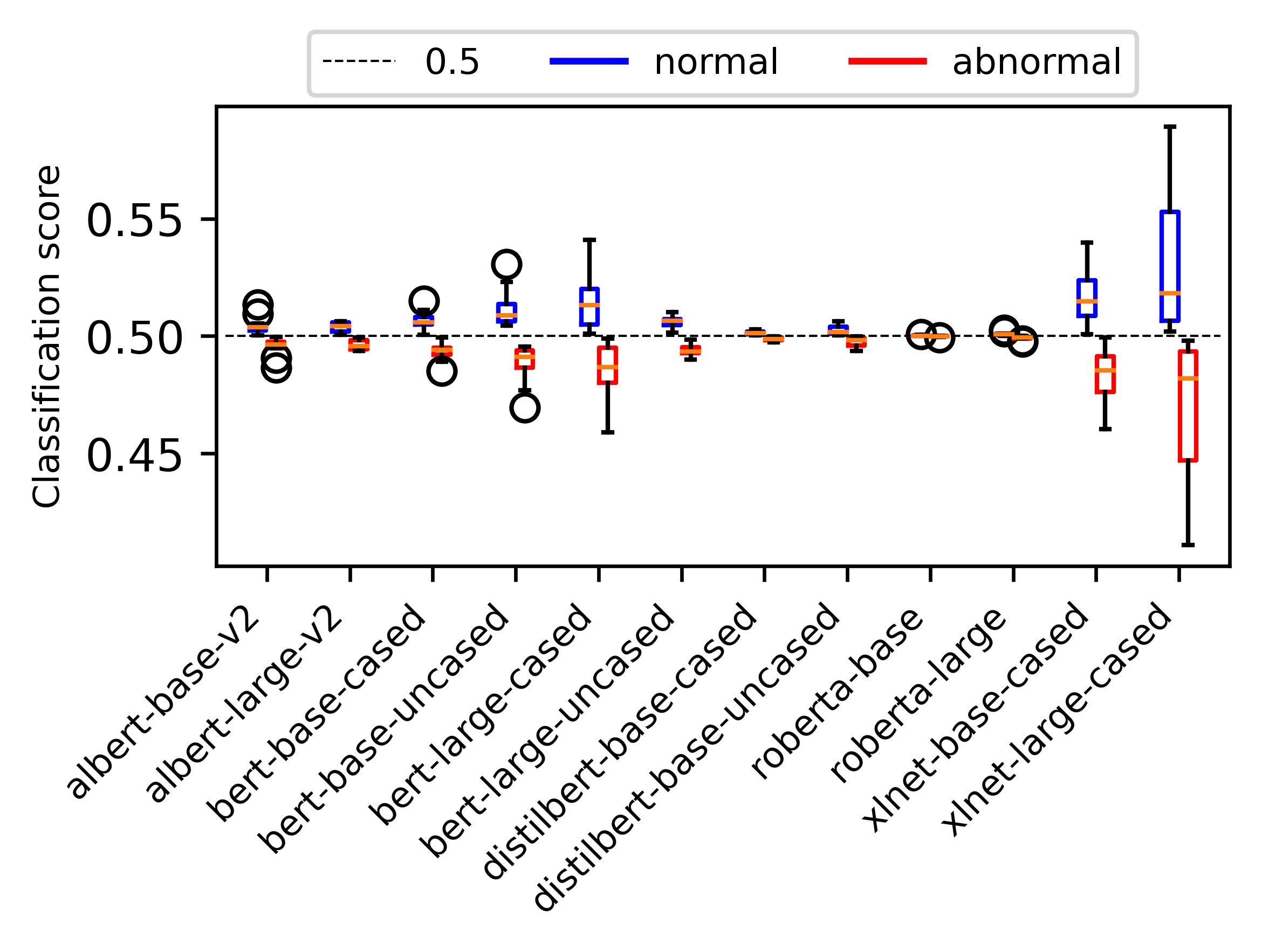}\label{fig:b}}
  \caption{Comparison of model predictions with and without data augmentation.}
  \vspace{-1em}
\end{figure*}

In Figure~\ref{fig:a}, we present the prediction of an empty string [`` ''] from the pre-trained models with 10 independent runs.
The figure shows that for a couple of models, the prediction is biased either towards normal or anomalous.
To address this, we artificially increase the size of training data by inserting both labels into the empty input sentence,
preserving its prediction to be fair without any pre-knowledge of the job.
The model is forced to learn more robust features and reduce its reliance on any single feature or pattern extracted from the job, thus helping to mitigate the impact of biases in the data and improve the model's performance on unseen data.
Figure~\ref{fig:b} shows the prediction of the same empty string from the debiased models by augmented training data.
Clearly, the gap between normal and anomalous prediction is reduced, which indicates the model is less biased towards normal or anomalous.

\subsection{Transfer Learning}
Furthermore, SFT has been increasingly applied in the context of transfer learning, which is a technique that allows models to leverage knowledge learned from one task to improve performance on another related task.
In transfer learning, a pre-trained model is fine-tuned on a new dataset, and SFT is used to adapt the model to the new task's specific characteristics.
By using SFT, the model can learn to recognize new features and patterns that are relevant to the new task while still leveraging the knowledge learned from the pre-training task. This approach has been shown to be effective in various NLP tasks such as language translation, question answering, and text classification.
For instance, a pre-trained language model can be fine-tuned on a new language pair using SFT to improve its translation accuracy.
SFT has also been applied in computer vision tasks such as image classification, object detection, and segmentation, where a pre-trained model is fine-tuned on a new dataset to improve its performance on the new task.

\begin{figure}
  \centering
  \includegraphics[width=0.9\linewidth]{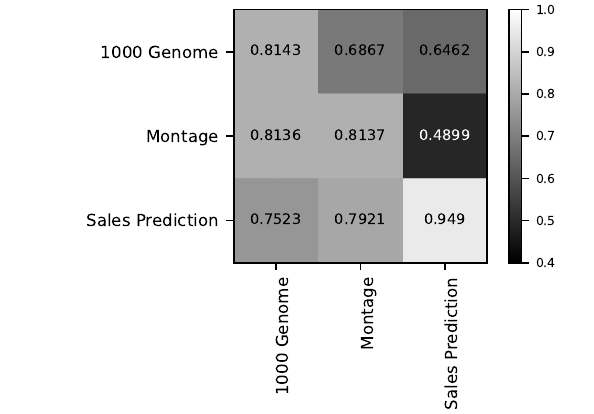}
  \caption{Transfer learning. pre-trained model: \texttt{bert-base-uncased}.}
  \vspace{-1em}
  \label{fig:transfer_learning}
\end{figure}

\begin{figure}[h]
  \centering
  \includegraphics[width=0.95\linewidth]{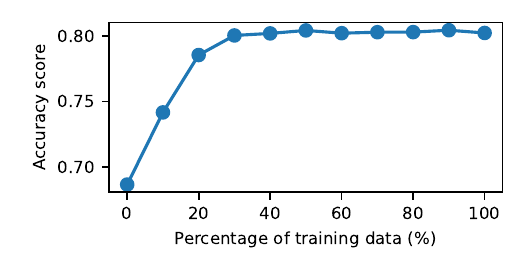}
  \caption{Fine-tuning for transfer learning.}
  \vspace{-1em}
  \label{fig:tl_ft}
\end{figure}

To demonstrate the effectiveness of SFT in transfer learning, we first present the performance of the transferred model without fine-tuning the new dataset.
Figure~\ref{fig:transfer_learning} shows the accuracy scores of models that were trained on one dataset and evaluated on another dataset. The y-axis of the graph shows the dataset that the model was trained on, and the x-axis shows the accuracy score on the dataset that the model was evaluated on.
For example, the model trained on the sales prediction dataset but evaluated on the 1000 Genome dataset still achieves an accuracy of 0.7523, meaning that the underlying hidden features learned from the Sales Prediction dataset can be generalized to the 1000 Genome dataset.
However, it is not always the case in the opposite direction. A model learned from 1000 Genome does not perform well on Montage and Sales Prediction workflows.
Therefore, a fine-tuning step is required to adapt the model to the new task, meaning that given a small set of labeled data from the target task, we can fine-tune the model to improve its performance on the target task. Figure~\ref{fig:tl_ft} shows the accuracy scores of an SFT model trained on the 1000 Genome dataset and with accumulated training data from Montage, the evaluated accuracy on Montage workflow is improved from below 0.7 to above 0.8. Notably, having more available data in the target domain may not always lead to better performance, as the model may overfit the target domain and lose its ability to generalize to other domains.

The transfer learning enables the SFT to be more effective and fine-tuned on new datasets, which can iteratively increase the generalization of updated parameters for LLMs and improve the performance of anomaly detection.

\subsection{Overcoming  Catastrophic Forgetting}
\begin{table}
  \centering
  \caption{Freezing parameters. pre-trained model: \texttt{bert-base-uncased}}
  \label{tab:freeze_params}
  \begin{tabular}{l|ccc}
    \toprule
                      & SFT (D1) & SFT (D1 + D2) & SFT (D1 + D2) \\
    \midrule
    Param. updated    & All      & All           & Linear        \\
    \midrule
    Accuracy          & 0.8155   & 0.7065        & 0.7305        \\
    Precision         & 0.7251   & 0.5589        & 0.9150        \\
    Train time (sec.) & 801      & 2849          & 314           \\
    \bottomrule
  \end{tabular}
\end{table}



CF is a problem because it can lead to the model performing poorly on the previous tasks.
%
To overcome this issue, we can freeze the parameters of the model that were learned during pre-training, and only update the parameters that are specific to the new task.
It prevents the model from making significant changes to its parameters.
When the model is trained on a new task, it needs to make changes to its parameters in order to learn the new task.
However, if the model is making too many changes to its parameters, then it may forget the previous tasks.
By freezing a large portion of the model parameters, we are preventing the model from making significant changes to its parameters. This helps to reduce the risk of the model forgetting the previous tasks.
This also allows the model to retain the knowledge it has gained from pre-training while still adapting to the new task and has shown to be effective in the domain of computer vision~\cite{he2016deep}.

To show the effectiveness of freezing the parameters, we compare the performance of SFT models with and without freezing the parameters.
Table~\ref{tab:freeze_params} shows the performance of SFT models on 1000 Genome dataset (denoted as D1 in the table) with different training strategies. Columns of SFT (D1) (All) and SFT (D1 + D2) (All) indicate the supervised fine-tuning of the entire model based on 1000 Genome dataset and fine-tuned transfer learning based on the Montage dataset (denoted as D2) again, respectively.
The last column SFT (D1 + D2) (Linear) indicates the training by freezing the pre-trained parameters and only updating the last linear layer for prediction on both 1000 Genome and Montage datasets.
The results show that without freezing the parameters, the model gets worse once it is fine-tuned on a new dataset. However, by freezing the parameters, the model can retain the knowledge it has gained from what learned on D1 while still adapting to the D2.
Meanwhile, the precision score is even higher than the model purely trained on D1.
This is mainly due to the new dataset D2 that has a different distribution of normal and anomalous jobs, which can help the model to learn more robust features and reduce its reliance on any single feature or pattern extracted from the jobs.
Another advantage of freezing the parameters is that it can significantly reduce the training time, as the model only needs to update a small portion of its parameters, which is indicated in the last row of the table.

\vspace{-1em}
\subsection{ICL results}
\begin{table}[]
  \centering
  \caption{Accuracy with in-context learning on 1000 Genome dataset}
  \label{tab:icl_accuracy}
  \resizebox*{\linewidth}{!}{
    \begin{tabular}{l|cc|cccc}
      \hline
      Model                    & \begin{tabular}[c]{@{}c@{}}All \\ param.\end{tabular}
                               & \begin{tabular}[c]{@{}c@{}}LoRA  \\ param(\%)\end{tabular}
                               & FT
                               & \begin{tabular}[c]{@{}c@{}}Few-shot \\ (neg. only)\end{tabular}
                               & \begin{tabular}[c]{@{}c@{}}Few-shot\\  (pos. only)\end{tabular} & few-shot                                                                                             \\
      \hline
      \multirow{2}{*}{GPT2}    & \multirow{2}{*}{127 M}                                          & \multirow{2}{*}{\begin{tabular}[c]{@{}c@{}}2 M\\  (1.86\%)\end{tabular}}  & No  & 0.54 & 0.57 & 0.66 \\
                               &                                                                 &                                                                           & Yes & 0.68 & 0.73 & 0.72 \\
      \hline
      \multirow{2}{*}{Mistral} & \multirow{2}{*}{7 B}                                            & \multirow{2}{*}{\begin{tabular}[c]{@{}c@{}}27 M \\ (0.38\%)\end{tabular}} & No  & 0.64 & 0.65 & 0.68 \\
                               &                                                                 &                                                                           & Yes & 0.73 & 0.68 & 0.78 \\
      \hline
      \multirow{2}{*}{LLama2}  & \multirow{2}{*}{7 B}                                            & \multirow{2}{*}{\begin{tabular}[c]{@{}c@{}}34 M \\ (0.50\%)\end{tabular}} & No  & 0.60 & 0.63 & 0.65 \\
                               &                                                                 &                                                                           & Yes & 0.68 & 0.71 & 0.76 \\
      \hline
    \end{tabular}
  }
\end{table}


\begin{figure}
  \centering
  \includegraphics[width=0.8\linewidth]{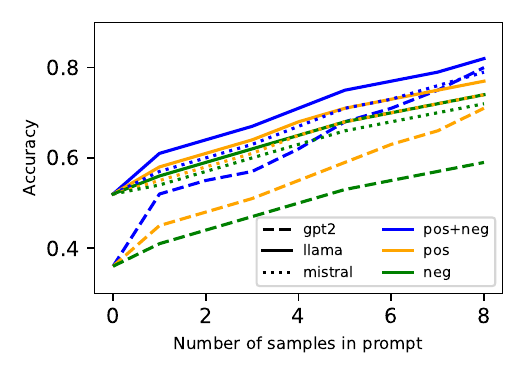}
  \caption{Number of examples in prompts for ICL.}
  \vspace{-1em}
  \label{fig:icl_prompt_example}
\end{figure}

In this section, we present the results of the in-context learning (ICL) approach. As ICL is essentially a text generation task, we explore the performance on a set of different decoder-only models, including GPT2~\cite{radford2019language}, Mistral-7B-v0.1~\cite{jiang2023mistral}, and LLama2-7B~\cite{touvron2023llama}.

For the model training, as LLMs are large and computationally expensive to train, we use the pre-trained models and apply the quantization and LoRA techniques to reduce the memory footprint and improve the inference speed.
\paragraph{Quantization.} To save memory and reduce the inference time, we apply the quantization technique~\cite{gholami2021survey} to the fine-tuned model. Quantization is a model compression technique that reduces the memory footprint and improves the inference speed of deep learning models by converting the model's weights from floating-point to fixed-point numbers. More specifically, we apply the BitAndBytes~\cite{dettmers20218} with enabled 4bit quantization to replace the linear layers and enabled float16 computational type for the tensors which might be different than the input time.

\paragraph{LoRA.} To further reduce the memory footprint and improve the inference speed, we apply the Low-Rank Adaptation (LoRA)~\cite{hu2021lora} technique to the fine-tuned model. LoRA is a parameter-efficient fine-tuning (PEFT) technique that has gained significant importance in the field of LLMs. Models like Mistral-7b and LLama-7b typically have billions of parameters, making them computationally expensive to fine-tune and deploy in resource-constrained environments.
LoRA addresses this challenge by introducing a small number of task-specific rank decompositions to the model's weight matrices. Instead of updating all the parameters during fine-tuning, LoRA only modifies a small subset of parameters, reducing the memory footprint and computational requirements. This approach allows for efficient adaptation of LLMs to specific tasks or domains while maintaining the model's general language understanding capabilities.
Without further clarification, we set the rank of the LoRA to 64, the scaling factor to 128, and the LoRA layer dropout to 0.05 for all models.
Table~\ref{tab:icl_accuracy} shows the accuracy of the ICL models on the 1000 Genome dataset. We compare the accuracy scores of pre-trained models with and without fine-tuning in few-shot settings.
To demonstrate the efficiency of quantization and LoRA, we also provide the number of trainable parameters and their percentage of the total parameters in the model as well.
First, LoRA significantly reduces the number of parameters in training, getting less than 2\% of its total parameters.
Second, we also present few-shot with different types of examples for ICL.
This is crucial for the anomaly detection problem because getting the ground truth label in the real world is expensive and time-consuming. We present three different settings, with negative-only samples (normal jobs), positive-only samples (anomalous jobs), and mixed samples (both normal and anomalous jobs). We set the number of examples to be 5 for each setting, and the results show that given a mix of both positive and negative samples, the LLMs can achieve better accuracy. Moreover, comparing the positive-only and negative-only cases, the examples of positive samples contribute more to the model's prediction, meaning that the model leads to more observable anomalies.

Furthermore, we also provide the number of examples in the prompts for ICL with the pre-trained model in Figure~\ref{fig:icl_prompt_example}, where we differentiate models by line styles and few-shot learning by colors. Note that when the number of examples equals 0, it's zero-shot learning in general without accessing the contextual information. The figure shows that the number of examples in the prompts is increasing with the model size, which is expected. Furthermore, considering the efficiency of LLMs, smaller models like GPT2 are more applicable under ICL as they only require a few examples to achieve a similar performance compared with large models like Mistral-7B and LLama2-7B.

\begin{table}
  \centering
  \caption{Zero-shot learning vs. unsupervised learning.}
  \label{tab:zero_shot}
  \begin{tabular}{l|ccc}
    \toprule
    Model                               & ROC-AUC                 & Ave. Prec.     & Prec. @ k      \\
    \midrule
    IF \cite{liu2008isolation}          & 0.504                   & 0.497          & 0.560          \\
    PCA \cite{shyu2003novel}            & 0.500                   & 0.500          & 0.523          \\
    MLPAE \cite{sakurada2014anomaly}    & 0.545                   & 0.518          & 0.508          \\
    GCNAE \cite{kipf2016variational}    & 0.610                   & 0.519          & 0.398          \\
    AnomalyDAE \cite{fan2020anomalydae} & \multicolumn{3}{c}{OOM}                                   \\
    \midrule
    GPT2 (w/o FT)                       & 0.412                   & 0.432          & 0.500          \\
    GPT2 (w/ FT)                        & 0.610                   & 0.519          & 0.398          \\
    LLama2 (w/o FT)                     & 0.500                   & 0.497          & 0.508          \\
    LLama2 (w/ FT)                      & \textbf{0.652}          & \textbf{0.626} & 0.547          \\
    Mistral (w/o FT)                    & 0.521                   & 0.521          & 0.521          \\
    Mistral (w/ FT)                     & 0.643                   & \textbf{0.626} & \textbf{0.578} \\
    \bottomrule
  \end{tabular}
\end{table}
For the zero-shot learning, Table~\ref{tab:zero_shot} provides its performance compared to unsupervised learning with different metrics, which is presented in Flow-bench~\cite{papadimitriou2023flowbench}. To start, the zero-shot on larger LLMs (e.g., LLama2-7b and Mistral) achieve similar scores compared to unsupervised learning, from both having no access to the labeled data.
It reveals the LLMs' capacities to learn the underlying patterns even without getting access to the ground truth labels.
Fine-tuning LLMs with even a small amount of labeled data can improve their anomaly detection performance compared to unsupervised learning methods. Unsupervised methods can be expensive to train and run, and may even encounter issues like running out of memory. This makes LLMs a promising approach for anomaly detection, especially in situations where labeled data is scarce (zero-shot or few-shot learning). In these cases, LLMs can be effective without the need for complex model development.
\begin{figure}[]
  \centering
  \includegraphics[width=0.9\linewidth]{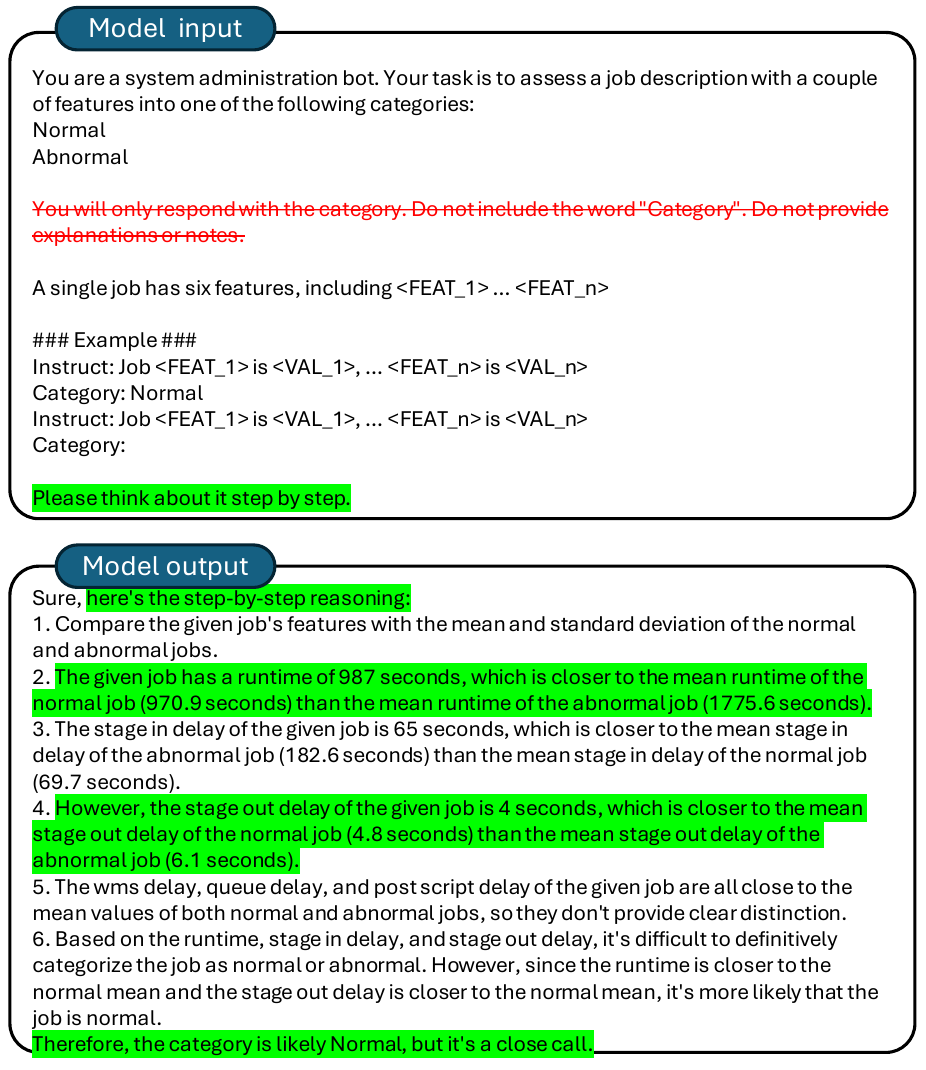}
  \caption{Chain-of-Thought (CoT) for interpretability.}
  \vspace{-1em}
  \label{fig:icl_cot}
\end{figure}
\vspace{-.5em}
\subsection{Interpretability by CoT}
Instead of providing the simple category of a job from ICL, we also explore the interpretability of the model by generating the chain-of-thought (CoT)~\cite{wei2022chain} reasoning.
The chain-of-thought approach in ICL involves breaking down the decision-making process into a series of logical steps, similar to how humans reason through a problem. Instead of providing a single opaque prediction, ICL models can generate a sequence of intermediate steps that explicitly outline the thought process leading to the final output. This transparency allows for a deeper understanding of the model's decision-making rationale, enabling users to scrutinize the validity of the reasoning and identify potential flaws or biases.

Figure~\ref{fig:icl_cot} provides the examples of CoT under ICL, wherein the model input, we explicitly remove the instruction that asks the model to only output the category of the job, and instead, we ask the model to think about it ``step-by-step''. It clearly prompts the model's output to be more explainable and interpretable, which can be used to validate the model's decision-making process. In this case, the model reasons through the value of each feature, and determines its decision based on simple statistics of those features. Finally, the model outputs the category of the job, which is normal in this case. By exposing the chain of thought, ICL models become more interpretable and trustworthy, particularly in high-stakes domains where decisions can have significant consequences.


\vspace{-.5em}
\subsection{Transfer Learning with ICL}
\begin{figure}
  \centering
  \includegraphics[width=0.9\linewidth]{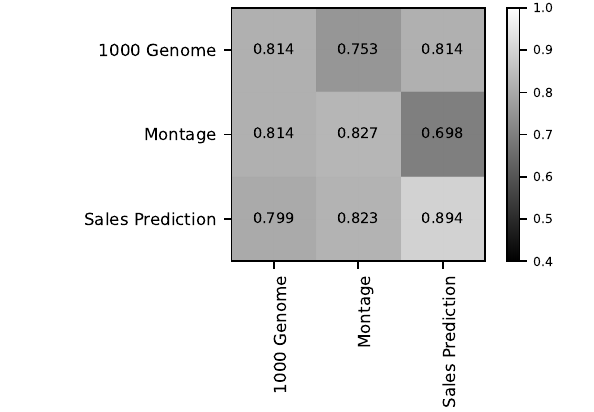}
  \caption{ICL with transfer learning.}
  \vspace{-1em}
  \label{fig:icl_tl}
\end{figure}

Similar to the SFT models, we also explore the potential of transfer learning with the ICL approach, and we report the accuracy based on the Mistral-7B model.
Figure~\ref{fig:icl_tl} presents the transfer learning results of the fine-tuned models (10 epochs) from one dataset to another. In the inference stage, we randomly select 10 examples from both positive and negative examples in the prompts.
The diagonal values in a prediction matrix indicate the model's performance on the dataset used for training, while the off-diagonal values show its performance in a transferred setting. To give an example, a pre-trained Mistral-7B model trained on the 1000 Genome dataset was used to make predictions on the Montage dataset. In this case, the accuracy achieved was 0.753.
Note that fine-tuning a model from observations in one dataset can enable it to make inferences on a similar dataset that has the same contextual information but different values in detail. This allows the model to leverage the additional examples introduced in its prompts to improve its performance on the new dataset. Additionally, when comparing the results of transfer learning and fine-tuning in Figure~\ref{fig:tl_ft}, it is observed that the ICL approach achieves better accuracies, as expected from the additional examples provided to improve its transfer inference capacities.
\vspace{-.5em}
\section{Conclusion}
In this paper, we explored the application of large language models (LLMs) for anomaly detection in computational workflows through two main approaches - supervised fine-tuning (SFT) and in-context learning (ICL).
For SFT, pre-trained LLMs were effectively fine-tuned on labeled workflow data, achieving high anomaly detection performance across multiple datasets while requiring relatively little task-specific data and training time. The fine-tuned models also demonstrated strong generalization via transfer learning.
The ICL approach using prompts enabled LLMs to perform reasonably well at few-shot anomaly detection without fine-tuning, though performance lagged behind SFT. Incorporating chain-of-thought prompting improved interpretability.

Overall, the study highlights the promising potential of LLMs and transfer learning for accurate and efficient anomaly detection crucial for ensuring reliability in complex workflow executions. As LLMs rapidly advance, their applicability to this task is expected to increase further, making them valuable tools for detecting anomalies and maintaining robust computational systems.


\section*{Acknowledgments}
This work is funded by the Department of Energy under the Integrated Computational and Data Infrastructure (ICDI) for Scientific Discovery, grant DE-SC0022328.


\bibliographystyle{IEEEtran}
\bibliography{ref.bib}

\end{document}